\documentclass[letterpaper]{article} 
\usepackage{array}
\usepackage{aaai23}  
\usepackage{times}  
\usepackage{helvet}  
\usepackage{courier}  
\usepackage[hyphens]{url}  
\usepackage{graphicx} 
\usepackage{natbib}  
\usepackage{caption} 
\usepackage{algorithm}
\usepackage{algorithmic}
\usepackage[dvipsnames, svgnames, x11names]{xcolor}
\usepackage{multirow}
\usepackage{booktabs}
\usepackage{amsmath}
\usepackage{cite}
\usepackage{subfig}
\usepackage{newfloat}
\usepackage{listings}
\DeclareCaptionStyle{ruled}{labelfont=normalfont,labelsep=colon,strut=off} 
\floatstyle{ruled}
\newfloat{listing}{tb}{lst}{}
\floatname{listing}{Listing}
\title{Unsupervised Cross-Domain Rumor Detection with Contrastive Learning and Cross-Attention}
\author{
    Hongyan Ran
    ,
    Caiyan Jia
    \thanks{Corresponding authors}
}
\affiliations{
    School of Computer and Information Technology \& Beijing Key Lab of Traffic Data Analysis and Mining\\ Beijing Jiaotong University, Beijing 100044, China\\
    
    \{hongyran,cyjia\}@bjtu.edu.cn
}

\usepackage{bibentry}

\begin{document}

\maketitle

\begin{abstract}
Massive rumors usually appear along with breaking news or trending topics, seriously hindering the truth. Existing rumor detection methods are mostly focused on the same domain, thus have poor performance in cross-domain scenarios due to domain shift. In this work, we propose an end-to-end instance-wise and prototype-wise contrastive learning model with cross-attention mechanism for cross-domain rumor detection. The model not only performs cross-domain feature alignment, but also enforces target samples to align with the corresponding prototypes of a given source domain. Since target labels in a target domain are unavailable, we use a clustering-based approach with carefully initialized centers by a batch of source domain samples to produce pseudo labels. Moreover, we use a cross-attention mechanism on a pair of source data and target data with the same labels to learn domain-invariant representations. Because the samples in a domain pair tend to express similar semantic patterns especially on the people's attitudes (e.g., supporting or denying) towards the same category of rumors,  the discrepancy between a pair of source domain and target domain will be decreased. We conduct experiments on four groups of cross-domain datasets and show that our proposed model achieves state-of-the-art performance. 
\end{abstract}

\section{Introduction}

Nowadays, with the rapid development of social media which has evolved into the primary source of news, more and more people are spending more time on social media platforms expressing what they see and hear, especially when it comes to hot topics 
or breaking events\footnote{https://user.guancha.cn/main/content?id=710205}. This creates a hotbed for the rumor subsisting and inspires the rumor publishers to make numerous rumors for their purposes. For instance, as rumors about the COVID-19 pandemic era spread rapidly, around 800 deaths, 5,000 hospitalizations, and 60 permanent injuries were recorded due to false claims that household bleach was an effective panacea for the virus \cite{coleman2020hundreds}. There are still many such rumors on social media, if not identified in time, sensational rumors may cause social panic during emergency events, and threaten the internet's credibility and trustworthiness. Thus, it has highly practical application value to detect rumors in an efficient way on social media platforms.
	
	\begin{figure}[t]
		\centering
		\includegraphics[width=1.0\columnwidth]{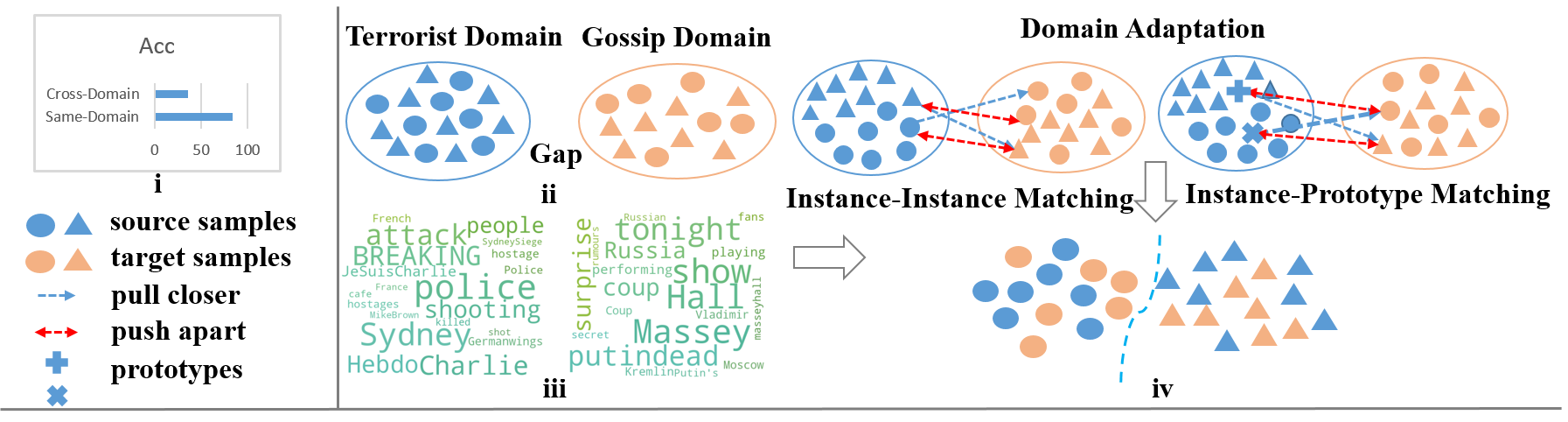} 
		\caption{An illustration example of unsupervised cross-domain rumor detection. 
  }
		\label{examples}
	\end{figure}
	
To solve the problem, various rumor detection methods have been proposed including traditional detection models and deep learning detection models. The traditional works mostly utilize hand-crafted features to perform rumor detection \cite{castillo2011information,kwon2013prominent,ma2015detect,jin2017detection}. The deep learning-based methods exploit the content of rumors to automatically detect rumors. These methods are developed from the original methods which view the textual information of source posts and user responses or user profiles as time series \cite{ma2016detecting,yu2017convolutional,ma2019detect,ma2021improving}, to the propagation structure-based methods which model rumor content and related responses as a tree structure to learn rumor representations \cite{RvNN,BiGCN,PLAN,PPA-WAE}, and finally to the multi-resource heterogeneous aggregation methods \cite{HGARD,MGAT-ESM,min2022divide} which achieve the best detection performance.
	
However, these methods mostly detect rumors under in-domain conditions. In practical scenarios, the real-world news platforms release various claims in different domains everyday, and newly emergent and time-critical domain events are difficult to acquire sufficient labeled data in time. If we directly use the verified posts of history domains to train these models and test them on the newly emergent domain data, we will get poor performance due to the domain shift. We take PPA-WAE model \cite{PPA-WAE} as an example to verify its performance on cross-domain settings. Since PHEME \cite{PHEME} dataset contains two domain events including `terrorist' events and `gossip' events, we view the two domain events as a cross-domain setting (See Figure \ref{examples}(\romannumeral2-\romannumeral3)). We compare the experimental results of PPA-WAE at the in-domain and the cross-domain settings which are shown in Figure \ref{examples}(\romannumeral1), and observe that the accuracy of the cross-domain is far below than that of the in-domain. It demonstrates that propagation structure-based state-of-the-art detection models while they perform well for the domain they are trained on (e.g., terrorist), perform poorly in other domains (e.g., gossip). The limited cross-domain effectiveness of methods to detect rumors is mostly due to the domain-specific word usage and the writing style of rumor content (See Figure \ref{examples}(\romannumeral3))
making the model biased toward the training domains.

To address these challenges, some works \cite{Low-Resource,mosallanezhad2022domain} propose domain-invariant feature learning algorithms for cross-domain rumor detection, while needing labeled target data as auxiliary information to train the models. Thus these models can not handle the settings where no labeled target data is available. Namely, these methods does not work at unsupervised cross-domain settings. 
 \cite{min2022divide} utilize an adversarial topic discriminator for topic agnostic feature learning to the unsupervised cross-domain rumor detection, but it has worse performance. Therefore, an effective method is needed to alleviate the challenges of unsupervised cross-domain rumor detection. Due to the consistency of the propagation patterns for rumors, no matter which domain they come from \cite{MGAT-ESM}, we hope that the method is able to pull closer to the same labeled samples and push apart the different labeled samples between the source domain and the target domain to decrease the domain gaps, so we believe that aligning the representation space of rumor-indicative patterns of different domains could adapt the features captured from the source data to that of the target data (See Figure 1(\romannumeral4)).
	
In this study, inspired by self-supervised contrastive learning \cite{he2020momentum,chen2020simple}, we propose an unsupervised cross-domain rumor detection method based on contrastive learning and cross-attention. Since the same categories are shared by both domains, we build instance-wise and prototype-wise contrastive learning to align features so as to reduce the discrepancy between source data and target data. It not only performs cross-domain samples feature alignment but also enforces the target samples to be aligned with the prototype of the corresponding source domain. 
Since target labels are not available, we use a clustering-based approach with carefully initialized centers on batch samples of a given source domain to produce pseudo labels. Moreover, we use a cross-attention mechanism on pairs of source data and target data with the same labels to learn domain-invariant features. Because these pairs express similar semantic patterns 
 learned on propagation paths of rumors, the discrepancy between domains will be decreased, thereby boosting the performance of our cross-domain rumor detection method.
 
The main contributions of this study are summarized in the following.
	\begin{itemize}
		\item We investigate the problem of cross-domain rumor detection and propose instance-wise and prototype-wise contrastive learning to align feature representations so as to reduce the discrepancy between domains.
		\item We construct a cross-attention mechanism between the source data and target data pairs with the same labels to learn domain-invariant features.
		\item We conduct experiments on four groups of cross-domain datasets and show that our proposed method achieves the best performance.
	\end{itemize}

	\section{Related Work}
	\subsection{Rumor Detection}
	Existing rumor detection methods mostly pay attention to the same domain rumor data and build various frameworks on it for well adapting to the tasks of rumor detection. For instance, sequence processing models leverage the textual contents from the source posts and user reply comments for rumor detection \cite{ma2016detecting,yu2017convolutional,ma2019detect,ma2021improving}, propagation structure-based methods \cite{RvNN,BiGCN,PLAN,PPA-WAE,liu2022nowhere,sun2022rumor} model the propagation paths as a tree attached with the textual content to build the semantics of posts and their propagation relationships, and some studies integrate the content of posts, relationships of user-post and user-user pairs, user profiles as a heterogeneous graph and have achieved the best performance for rumor detection \cite{HGARD,li2021joint,MGAT-ESM}. Lately, some researchers study semi-supervised cross-domain rumor detection using supervised contrastive learning and reinforcement learning \cite{Low-Resource,mosallanezhad2022domain}. Different from these researches, we mainly focus on unsupervised cross-domain rumor detection, where the labels of the target domain are unavailable. Recently, \cite{min2022divide} propose an unsupervised cross-domain rumor detection method using an adversarial topic discriminator for topic-invariant feature learning with limited performance. In addition, the model demands complex multi-source information as inputs, whereas our model only considers rumor content and its social text.

	\subsection{Unsupervised Cross-Domain}
	Unsupervised cross-domain (UCD) aims to learn a model that is able to achieve good classification accuracy without any annotation in a target domain \cite{ben2010theory}. Existing UCD methods mainly appear in the field of computer vision which includes domain-level and category-level apporaches. Domain-level approaches use the Maximum Mean Discrepancy (MMD) to mitigate the distribution divergence between the source and target domains by pulling them into the same distribution at different scale levels \cite{ganin2015unsupervised,long2018conditional}.
	Category-level methods align each category distribution between the source domain and target domain by pushing the target samples to the distribution of source samples in each category
	\cite{saito2018maximum,du2021cross,xu2021cdtrans}. Recently, UCD has been applied to various applications such as text classification \cite{zou2021unsupervised,li2022domain} and sentiment 
 analysis \cite{du2020adversarial,ghosal2020kingdom}, etc.
	In this work, we intent to introduce the category-level feature alignment and domain-invariant feature learning between domains into cross-domain rumor detection tasks.

	\subsection{Self-Supervised Contrastive Learning}
	The core idea of self-supervised contrastive learning is to learn from positive samples and benefit from correcting negative ones, which has been successfully applied to many fields. In computer vision, a large collection of works \cite{he2020momentum,chen2020simple} learn self-supervised image representation by minimizing the distance between two views of the same image.
	In natural language processing, 
	studies suggest that contrastive learning is promising in the semantic textual similarity \cite{gao2021simcse}, stance detection, and short text clustering \cite{mohtarami2019contrastive,zhang2021supporting}. In addition, contrastive learning has successfully promoted the development of representation learning of graph-structured data \cite{qiu2020gcc,you2020graph,zhu2021graph}.
	Our work is inspired by self-supervised contrastive learning, but the difference is that we use supervised contrastive learning for unsupervised cross-domain rumor detection tasks.
	
	\begin{figure*}[t]
		\centering
		 \includegraphics[width=0.87\textwidth,height=0.39\textwidth]{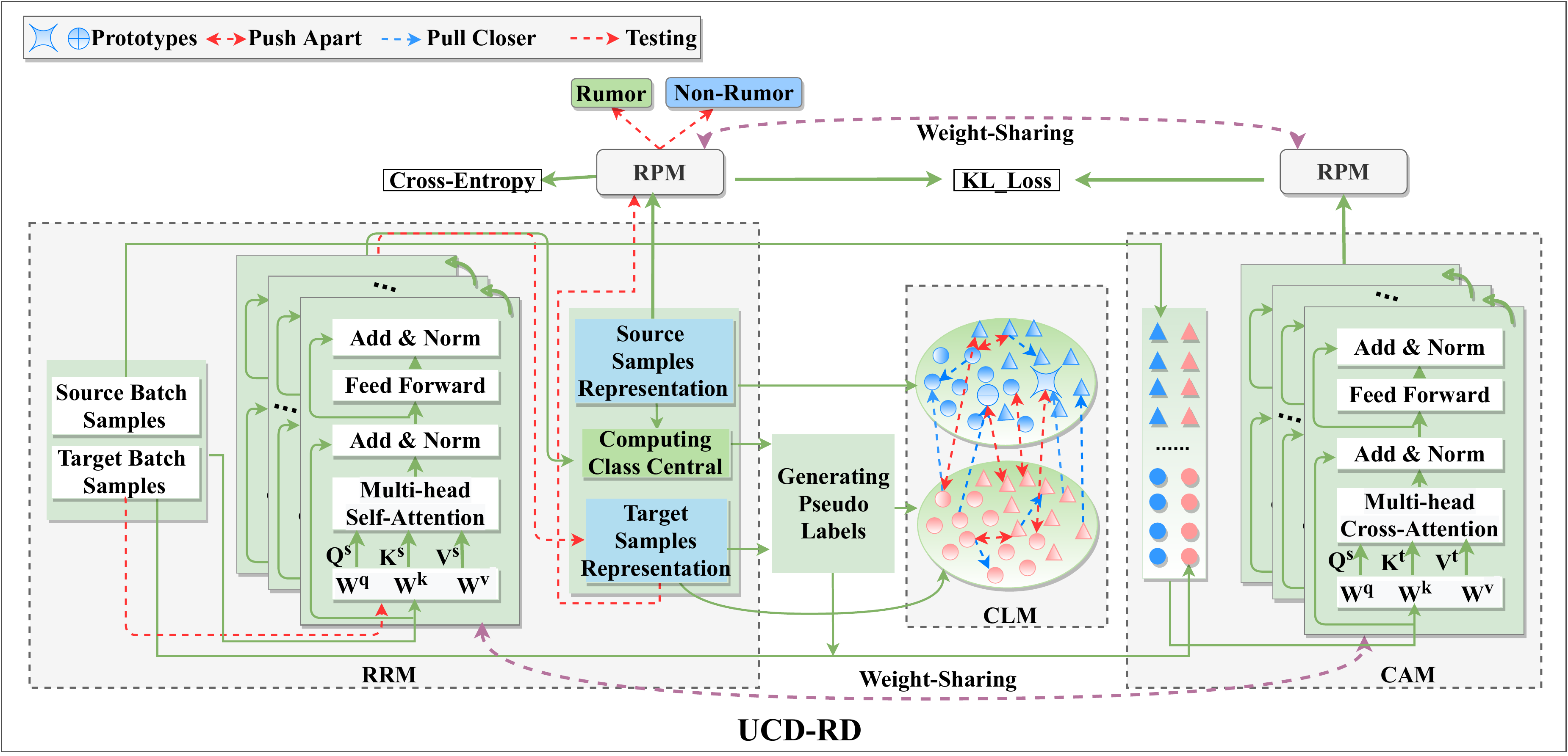} 
		\caption{The proposed UCD-RD framework. It consists of four main components including RRM, CLM, CAM and RPM. 
        }
	\label{fig2}
		
	\end{figure*}

	\section{Problem Statement}
	\subsubsection{Definition} 
 Unsupervised cross-domain rumor detection aims to transfer models learned from a labeled source data to an unlabeled target data. 
	Given a labeled rumor dataste $D^s=\{({C_i}^s,{y_i}^s)\},i=1,2,3...,n_s$ from the source domain, where ${C_i}^s$ denotes the set of post and comment contents described by ${C_i}^s=\{{s_i}^s,{r_{i1}}^s,...,{r_{i|i|-1}}^s\}$, in which ${s_i}^s$ is a source tweet, and ${r_{ij}}^s$ is the $j$-th comment text and $|i|$ refers to the number of source post and comments in ${C_i}^s$,
	${y_i}^s \in y^s$ denotes the corresponding label. 
	In the target domain, also given an unlabeled dataset $D^t=\{{C_i}^t\}, i=1,2,...,n_t$, where ${C_i}^t$ represents the rumor of $i$-th unlabeled sample in the target domain and also includes a series of reply posts.
	Our goal is to predict labels of testing samples in the target domain using a model $f_t:C^t\rightarrow y^t$ trained on $D^s\cup D^t$, where the target label space $y^t$ is equal to the source label space $y^s$.
	
	\subsubsection{Data processing} We build each rumor $C_i$ whichever comes from the source domain or the target domain as a propagation tree.
	Since the propagation paths aggregated method \cite{PPA-WAE} can effectively acquire the representation of a rumor, we construct the rumor $C_i$ as a set of propagation paths $P_i=\{P_{i1},P_{i2},...,P_{i|i|}\}$, where $|i|$ denotes the propagation path number of the rumor $C_i$, $P_{ij}$ represents $j$-th path of the rumor corresponds to the path from source tweets $s_i$ to the $j$-th leaf node, and it can be represented by $P_{ij}=[s_i,n_{ij1},n_{ij1},...,n_{ij|n|}]$, where $n_{ijk}$ is the $k$-th node of the $j$-th path in the $i$-th rumor propagation tree, $|n|$ is the number of nodes along the path, and each node includes a series of words. 
	We use GloVe 300d word embedding vectors \cite{glove} to initialize each word in a propagation path $P_{ij}$.
	
	\section{Proposed Model}
	In this section, we will describe our proposed model, {\bf{U}}nsupervised {\bf{C}}ross-{\bf{D}}omain {\bf{R}}umor {\bf{D}}etection with Contrastive Learning and Cross-Attention ({\bf{UCD-RD}}). As shown in Figure \ref{fig2}, the {\bf{UCD-RD}} model has four main components: Rumor Representation Module (RRM) , Contrastive Learning Module (CLM), Cross-Attention Module (CAM) and Rumor Prediction Module (RPM). 
	
	\subsection{Rumor Representation Module (RRM)}
	Since the self-attention mechanism in the transformer network \cite{vaswani2017attention} enables the model to effectively model long-range dependencies,
	hence we use the transformer network and its Multi-head Attention (MHA) module to learn the propagation structure information to obtain a good rumor representation.
	
	Given a set of propagation path $P_i$ of rumor $C_i$ which comes from the source domain or the target domain, we apply max-pooling to each path $P_{ij} \in P_i$ in the linear structure to obtain it's path representation $X_{P_{ij}}$, the sequence of the path embedding for the rumor $C_i$ can be represented as $X_{P_i}=(X_{P_{i1}},X_{P_{i2}},...,X_{P_{i|i|}})$.
	
	Next, we use the MHA module to learn path propagation embedding for each rumor. In detail, an MHA layer is made up of a self-attention layer and fully connected feed-forward layer. The scaled dot-product self-attention layer is the core component of the transformer. It uses an attention score to consider the relation between inputs. The inputs consist of queries and keys of dimension $d_k$, and values of dimension $d_v$. The dot products of a query (representing a specific path $X_{P_{ij}} \in X_{P_i}$) with all keys and apply a softmax function to obtain the attention score on the values. The 
	set of queries and those of the keys and the values are packed together into matrices $Q$, $K$, and $V$.
	For each head $j$, these input matrices will be projected into $d^k$, $d^k$, $d^v$ dimension subspaces as $Q_j$, $K_j$, $V_j$ through trainable linear projections $W_{j}^Q\in R^{d\times d^k}$, $W_{j}^K\in R^{d\times d^k}$, $W_{j}^V\in R^{d\times d^v}$, respectively. We denote the $j$-th head by $head_j$ as follows:
	\begin{equation}
	head_j=f_{att}(Q_j,K_j,V_j)=softmax(\frac{Q_jK_{j}^T}{\sqrt{d^k}})V_j
	\end{equation}
	
	For the rumor $C_i$, the final output of MHA can be calculated as a linear projection of concatenation of $h$ heads:
	\begin{equation}
	O_i=f_{MHA}=concat(head_1,...,head_h)W^o
	\end{equation}
	where $W^o\in R^{hd^v\times d}$.
	
	The output $O_i$ of the self-attention layer is then passed through a fully connected feed-forward layer consisting of two linear units with the Relu activation to acquire all the path representations of the rumor $C_i$. And finally, we use the max-pooling operation to aggregate the path embedding to obtain the representation $\hat{O}_i$ of the rumor $C_i$.

	\subsection{Contrastive Learning Module (CLM)}
	Inspired by the success of self-supervised contrastive learning \cite{he2020momentum,chen2020simple}, we model cross-domain rumor detection using instance-wise and prototype-wise contrastive learning to automatically learn how to align the sample representations from both labeled source domain $D^s$ and unlabeled target domain $D^t$, and use in-domain and cross-domain two aspects to perform this procedure. The core idea is to make the representations of source data and target data from the same class closer while keeping representations from different classes far away. 
	As the labels of the target data $D^t$ are unavailable, we propose a clustering-based approach with initialized centers by batch samples of source data $D^s$ to produce pseudo labels.
	
	\subsubsection{In-domain Contrastive Learning.} Given a batch of rumor samples from the source domain $D^s$, 
	we firstly obtain the representation for these samples according to the RRM. To make rumor representation in the source domain more discriminative, we then use an instance-wise contrastive learning objective to cluster the same class and separate different classes of samples, the objective function is computed as:
	\begin{equation}
	\begin{aligned}
	L_{SCL}^s = -\frac{1}{|B^s|}\sum_{i=1}^{|B^s|}\sum_{j=1}^{|B^s|}{\bf{1}}_{y_{i}^s=y_{j}^s}
\\log\frac{exp(sim(\hat{O}_{i}^s,\hat{O}_{j}^s)/\tau)}{\sum_{k=1}^{|B^s|}exp(sim(\hat{O}_{i}^s,\hat{O}_{k}^s)/\tau)}
	\end{aligned}
	\end{equation}
	where $|B^s|$ is the size of a source domain batch, $\bf{1}$ is an indicator. $sim(\cdot)$ denotes the cosine similarity function and $\tau$ controls the temperature.
	
	For the target data, the ground-truth labels are unavailable, so we exploit the cluster-based method to produce the pseudo labels for them (as will be introduced below), and then we also use instance-wise contrastive learning to pull closer to the same label samples and push apart the different label samples. Therefore, we can use a batch of target samples to compute $L_{SCL}^t$. The overall loss for in-domain is combined the two losses with different proportions $\alpha_i$:
	\begin{equation}
	L_{ICL} =\alpha_1L_{SCL}^s + \alpha_2L_{SCL}^t, s.t. \sum\nolimits_{i}{\alpha_i}=1
	\end{equation}

	\subsubsection{Cross-Domain Contrastive Learning.} We now introduce how to form pairs to learn domain invariant features with contrastive learning. Since samples from the source domain and target domain belong to the same set of classes, we build upon this assumption to reduce domain shift. More specifically, we hypothesize that samples within the same category are close to each other while samples from different classes lie far apart, regardless of which domain they come from. More formally, given an anchor sample in the target domain, and it forms a positive pair with a sample in the same class from the source domain, we formulate the instance-wise cross-domain contrastive loss as:
	\begin{equation}
	\begin{aligned}
	L_{CCL}^{t\rightarrow s} = -\frac{1}{|B^t|}\sum_{i=1}^{|B^t|}\sum_{j=1}^{|B^s|}{\bf{1}}_{y_{i}^t=y_{j}^s}\\log\frac{exp(sim(\hat{O}_{i}^t,\hat{O}_{j}^s)/\tau)}{\sum_{k=1}^{|B^s|}exp(sim(\hat{O}_{i}^t,\hat{O}_{k}^s)/\tau)}
	\end{aligned}
	\end{equation}
	
	The cross-domain loss forces intra-class distance to be smaller than inter-class distance for samples from different domains so as to reduce domain shift. 
	Alternatively, we can also use source samples as anchors and compute $L_{CCL}^{s\rightarrow t}$ loss.
	
	In order to explicitly enforce learning domain-aligned
	and more discriminative features in both source and target domains, we perform cross-domain prototype-wise contrastive learning. Our method discovers positive matching as well as negative matchings between instance and cluster prototypes from the target domain to the source domain. Specifically, given a feature vector $\hat{O}_{i}^t$ in the target domain, and prototypes $\{cen_{j}^s\}_{j=1}^{N_c}$ 
	which denote sample averages with the same label in the source batch data,
	we use prototype-wise contrastive learning as follows to train our model.
	\begin{equation}
	\begin{aligned}
	L_{Pro}^{t\rightarrow s} = -\frac{1}{|B^t|}\sum_{i=1}^{|B^t|}\sum_{j=1}^{N_c}{\bf{1}}_{y_{i}^t=y_j}\\log\frac{exp(sim(\hat{O}_{i}^t,cen_{j}^s)/\tau)}{\sum_{k=1}^{N_c}exp(sim(\hat{O}_{i}^t,cen_{k}^s)/\tau)}
	\end{aligned}
	\end{equation}
	where $N_c$ is the number of classes, $|B^t|$ is the size of a target domain batch. 
	
	The cross-domain contrastive loss can be denoted as:
	\begin{equation}
	L_{CCL} =L_{CCL}^{t\rightarrow s} + L_{CCL}^{s\rightarrow t} + L_{Pro}^{t\rightarrow s}
	\end{equation}
	
	Therefore, the total loss function of the CLM can be represented as follows:
	\begin{equation}
	L_{CL} =\beta_1L_{ICL} +\beta_2L_{CCL}, s.t. \sum\nolimits_{i}{\beta_i}=1
	\end{equation}

	\subsubsection{Pseudo Labels for a Target Domain.} The ground-truth labels from a target domain $D^t$ are unavailable during training, and thus we leverage k-means clustering \cite{kang2019contrastive} to produce pseudo labels.
	Since K-means is sensitive to initialization, the correspondence is unknown between using randomly generated clusters and predefined categories.
	To mitigate this issue, we set the number of clusters to be the number of classes $N_c$ and use class prototypes of batch samples from the source domain $D^s$ as initial clusters. Formally,  we first compute the centroid of each category using the sample average in the source batch and the initial cluster center $cen_{m}^t$ for the $m$-th class is defined as:
	\begin{equation}
	cen_{m}^t=cen_{m}^s=\frac{1}{|B_m^s|}\sum_{i=1}^{|B^s|}{\bf{1}}_m\hat{O}_i^s
	\end{equation}
	where $|B_m^s|$ is the sample numbers of $m$-th class in the batch $B^s$. Given a batch of features from the target domain, we perform K-means clustering using these initialized centers. 
	Once clustering is finished, each sample in the target domain $C_{i}^t$ is associated with a pseudo label $y_{i}^t$.
	
	\begin{table*}[!htp]
             \normalsize
		\centering

		\begin{tabular}{p{2.73cm}<{\centering}|p{0.85cm}<{\centering}p{0.9cm}<{\centering}|p{1cm}<{\centering}p{1.3cm}<{\centering}|p{0.6cm}<{\centering}p{2.65cm}<{\centering}|p{0.6cm}<{\centering}p{2.5cm}<{\centering}}
			\toprule
			\multirow{2}[2]{*}{Statistics} & \textit{source} & \textit{target} & \textit{source} & \textit{target} & \textit{source } & \textit{target} & \textit{source} & \textit{target} \\
			& \textit{Terrorist} & \textit{Gossip} & \textit{Twitter15} & \textit{Twitter16} & \textit{Twitter} & \textit{Twitter-COVID19} & \textit{Weibo} & \textit{Weibo-COVID19} \\
			\midrule
			\midrule
			\# of claims & 5940  & 485   & 1490  & 818   & 1154  & 400   & 4649  & 399 \\
			\midrule
			\# of tree nodes & 83,860 & 15,225 & 41,266 & 19,376 & 60,409 & 406,185 & 1,956,449 & 26,687 \\
			\midrule
			\# of non-rumors & 3907  & 116   & 374   & 205   & 579   & 148   & 2336  & 146 \\
			\midrule
			\# of rumors & 2033  & 369   & 1116  & 613   & 575   & 252   & 2313  & 253 \\
			\midrule
			Avg. \# of posts/tree & 15    & 32    & 28    & 24    & 52    & 1015  & 420   & 67 \\
			\bottomrule
		\end{tabular}%
            \caption{Statistics of the cross-domain datasets}
		\label{tab1}%
	\end{table*}%
	
	\subsection{Cross-Attention Module (CAM)}
	In the previous section, we utilize self-attention in RRM for samples both in source and target domains for learning the feature representations. Since the cross-attention 
 has been proved robust to the noisy input pairs for better feature alignment \cite{xu2021cdtrans}, owing to the noise of the pseudo labels in the target domain, we use the cross-attention mechanism to robust our model. Since each propagation path of rumors with the same labels represents similar semantic patterns, 
	especially on the people's attitudes (e.g., supporting or denying),
	thus we perform the cross-attention for rumor pairs on cross-domain samples with the same labels learning domain-invariant features. Such design explicitly enforces the framework to learn discriminative domain-specific and domain-invariant representations simultaneously according to self-attention and cross-attention.
	
	The CAM is derived from the self-attention module. The difference is that the input of cross-attention is a pair of rumors with the same labels which come from the cross-domain data, $i.e. X_{P_i}^s$ and $X_{P_j}^t$.
	Its query and key/value are from $X_{P_i}^s$ and $X_{P_j}^t$ respectively. The cross-attention score can be calculated as follows:
	\begin{equation}
	f_{att_{cross}}(Q^s,K^t,V^t)=softmax(\frac{Q^s(K^{t})^T}{\sqrt{d^k}})V^t
	\end{equation}
	where $Q^s$ are quries from $X_{P_i}^s$, and $K^t$, $V^t$ are keys and values from $X_{P_j}^t$. 
	For each output, it is calculated by multiplying $V^t$ with attention weights, which comes from the similarity between the corresponding query in $X_{P_i}^s$ and all the keys in $X_{P_j}^t$. As a result, among all paths in $X_{P_j}^t$, the path that is more similar to the query of $X_{P_i}^s$ would hold a larger weight and contribute more to the output. In other words, the output of the CAM manages to aggregate the two input rumors based on their similar paths.
	
	The CAM not only aligns distributions of two domains but is robust to the noise in the input pairs thanks to the cross-attention mechanism. Thus we use the output of the CAM to guide the model's training.
	We minimize the KL-divergence between predictions, which are computed by the Rumor Prediction Module (RPM) using the outputs of cross-attention and self-attention.
	\begin{equation}
	\begin{aligned}
	p_{cross_i}={\bf{RPM}}(\hat{O}_{cross_i}^s),
	p_{i}={\bf{RPM}}(\hat{O}_{i}^s)
	\end{aligned}
	\end{equation}
	\begin{equation}
	L_{CA}^s=KL^s(p_{cross}||p) =\sum_{i=1}^{|B^s|}p_{cross_i}(log\frac{p_{cross_i}}{p_i})
	\end{equation}
	where $\hat{O}_{cross_i}^s$ is the output of the CAM.

	\begin{table*}[t]
		\centering
            

		\begin{tabular}{p{1.86cm}<{\centering}|p{0.6cm}<{\centering}p{0.6cm}<{\centering}p{0.8cm}<{\centering}|p{1cm}<{\centering}p{1cm}<{\centering}p{1cm}<{\centering}|p{0.6cm}<{\centering}p{0.8cm}<{\centering}p{0.8cm}<{\centering}|ccc}
			\toprule
			\multirow{2}[4]{*}{Methods} & \multicolumn{3}{c|}{\textit{Terrorist$\rightarrow$Gossip}} & \multicolumn{3}{c|}{\textit{Twitter$\rightarrow$Twitter\_COVID19}} & \multicolumn{3}{c|}{\textit{Twitter15$\rightarrow$Twitter16}} & \multicolumn{3}{c}{\textit{Weibo$\rightarrow$Weibo\_COVID19}} \\
			\cmidrule{2-13}          & Acc.   & N.($F_1$)     & R.($F_1$)     & Acc   & N.($F_1$)     & R.($F_1$)     & Acc.   & N.($F_1$)     & R.($F_1$)     & Acc.   & N.($F_1$)     & R.($F_1$) \\
			\midrule
			\midrule
			LSTM  & 33.08 & 34.56 & 30.31 & 41.23 & 33.96 & 42.58 & 60.78 & 46.15 & 63.25 & 41.57 & 40.25 & 42.79\\
			CNN & 32.57 & 35.62 & 29.68 & 40.58    & 28.47 & 44.95 & 61.25 & 47.32 &  62.85 & 42.09 & 38.21 & 43.83 \\
			Rumor-GAN & 32.54 & 35.48 & 29.37 & 41.96 & 35.68 & 43.12 & 63.24 & 45.27 & 65.38 & 43.23 & 39.61 & 45.83 \\
			\midrule
			TD-RvNN & 33.59 & 32.47 & 28.16 & 43.55 & 40.09 & 45.78 & 69.77 & 47.96 & 71.28 & 47.90  & 43.66 & 54.78 \\
			BU-RvNN & 32.68 & 30.12 & 21.56 & 41.33 & 38.58 & 42.25 & 68.47 & 42.36 & 69.28 & 45.18 & 38.76 & 50.53 \\
			Bi-GCN & 34.84 & 31.04 & 27.41 & 51.67 & 31.72 & 46.37 & 75.15 & 42.23 & \underline{83.17} & \underline{61.23} & 44.08 & 68.11 \\
			PLAN  & 30.79 & 40.94 & 16.40  & 45.50  & \textbf{47.60} & 43.23 & 75.04 & 56.92 & 82.43 & 38.44 & \textbf{46.06} & 28.33 \\
			PPA-WAE & 41.56 & 38.33 & 45.79 & 46.79 & 43.28 & 48.46 & 74.89 & 60.08 & 77.69 & 57.56 & 35.33 & 65.26 \\
			\midrule
			UCD-CEloss & \underline{74.01} & \underline{66.49} & \underline{75.97} & \underline{58.87} & 33.99 & \underline{67.78} & \underline{75.15} & \underline{60.58} & 81.08 & 57.96 & 41.60 & \underline{68.36} \\
			{\bf{UCD-RD}} & \textbf{84.88} & \textbf{83.04} & \textbf{86.35} & \textbf{66.50} & \underline{45.34} & \textbf{76.72} & \textbf{79.47} & \textbf{63.53} & \textbf{85.72} & \textbf{68.92} & \underline{45.13} & \textbf{78.32} \\
			
			{\bf{$\uparrow$ (\%)}} & \textbf{14.69} & \textbf{24.92} & \textbf{13.66} & \textbf{12.96} & -4.75 & \textbf{13.19} & \textbf{5.75} & \textbf{4.87} & \textbf{3.07} & \textbf{12.56} & -2.02 & \textbf{14.57} \\
			\bottomrule
		\end{tabular}%
            \caption{Rumor detection results (\%) on four groups of cross-domain datasets (N: Non-Rumor; R: Rumor)}
		\label{tab2}%
	\end{table*}%

	\subsection{Rumor Prediction Module (RPM)}
	The RPM consists of a fully collected layer with softmax. Since the labels of the source domain samples are available, we exploit the RPM to acquire their prediction and then use the cross-entropy loss function to train the RPM and RRM.
	\begin{equation}
	L_{CE}^s = -\frac{1}{|B^s|}\sum_{i=1}^{|B^s|}y_{i}^slog(p_i)
	\end{equation}
	\begin{equation}
	p_i={\bf{RPM}}(\hat{O_i}^s)=softmax(FC(\hat{O_i}^s))
	\end{equation}
	
	We combine all of the loss functions together to jointly train our model, and can be represented as follows:
	\begin{equation}
	Loss=\gamma_1L_{CE}^s + \gamma_2L_{CL} + \gamma_3L_{CA}^s, s.t. \sum\nolimits_{i}{\gamma_i}=1
	\end{equation}
	where $\gamma_i$ is hyperparameters. Once the model is well trained, the unlabeled target data first adopts the RRM to acquire its vector representation, then passes through the RPM to obtain the label distribution according to Equ.14, and finally uses the maximal assignment to the $N_c$ values to get the final label for each rumor.

	\section{Experiments}
	\subsection{Datasets and Settings}
	\subsubsection{Datasets} We evaluate the UCD-RD model on four groups of real-world cross-domain rumor datasets. 
	The first group of 
	data comes from PHEME \cite{PHEME} dataset which includes terrorist domain and gossip domain, 
	more details
	are listed in Supplementary Material\footnote{https://github.com/rhy1111/Supplementary\_Material}. 
	The second group of cross-domain data is Twitter dataset \cite{ma2017detect} and Twitter-Covid19 dataset \cite{Low-Resource}. The third group of datasets includes the Twitter15 dataset and the Twitter16 dataset \cite{RvNN}. The fourth group of cross-domain data is the Chinese Weibo dataset \cite{ma2016detecting} and the Weibo-Covid19 dataset \cite{Low-Resource}. 
	These cross-domain datasets contain two binary labels: Non-Rumor (N) and Rumor (R). The statistics of the four groups of cross-domain datasets are shown in Table \ref{tab1}.
	\subsubsection{Experimental Setup} We compare the UCD-RD method with some state-of-the-art baselines
 including:
	\begin{itemize}
		\item {\bf{LSTM}} \cite{ma2016detecting} is an LSTM-based rumor detection model to learn feature representations of relevant posts over time.
		\item {\bf{CNN}} \cite{yu2017convolutional} uses a 
		CNN model for misinformation identification by framing the relevant posts as a fixed-length sequence. 
		\item {\bf{Rumor-GAN}} \cite{ma2019detect} uses a
		 generative adversarial network (GAN) in which a generator is designed to produce conﬂicting voices to pressurize the discriminator to learn stronger rumor representations.
		\item {\bf{TD-RvNN}} \cite{RvNN} exploits a top-down tree-structured recursive neural network for learning the propagation of rumors.
		\item {\bf{BU-RvNN}} \cite{RvNN} utilizes a bottom-up tree-structured recursive neural network for learning the propagation of rumors.
		\item {\bf{Bi-GCN}} \cite{BiGCN} is a GCN-based model based on conversation trees to learn rumor representations.
		\item {{\bf{PLAN}}} \cite{PLAN} uses a transformer-based model for rumor detection to capture long-distance interactions between any pair of involved tweets.
		\item {\bf{PPA-WAE}} \cite{PPA-WAE} utilizes a neural topic model which is combined with a feed-forward network on propagation
		trees to learn the semantics of the trees and their propagation patterns.
		\item {\bf{UCU-CEloss}} is a variant of the UCD-RD model which only uses the $L_{CE}$ loss to train the model.
	\end{itemize}
	
	We implement LSTM and CNN models with Keras\footnote{https://keras.io/}, other baseline methods and our model with Pytorch\footnote{https://pytorch.org/}. For these cross-domain datasets, we evaluate the Accuracy (Acc.), F1 measure ($F_1$) on each class. The dimension of each rumor hidden feature vector is 300. The training process is iterated upon 300 epochs. The temperature $\tau$ is 0.1.
	
	\begin{table*}[htbp]
            \normalsize
		\centering

		\begin{tabular}{c|c|c|rrrrrrr}
			\toprule
			\multicolumn{3}{c|}{} & \multicolumn{1}{l}{$L_{CE}^s$} & \multicolumn{1}{l}{+$L_{SCL}^s$} & \multicolumn{1}{l}{+$L_{SCL}^t$} &
			\multicolumn{1}{l}{+$L_{CDC}^{t\rightarrow s}$} & \multicolumn{1}{l}{+$L_{CDC}^{s\rightarrow t}$} &  \multicolumn{1}{l}{+$L_{Pro}^{t\rightarrow s}$} & \multicolumn{1}{l}{+$L_{CA}$}  \\
			\midrule
			\midrule
			\multicolumn{3}{c|}{{\textit{Terrorist$\rightarrow$Gossip}}} & 74.01      & 74.58      &  75.65     &  77.90     &  79.01     &  79.58     &  {\bf{84.88}}   \\
			\midrule
			\multicolumn{3}{c|}{{\textit{Twitter$\rightarrow$Twitter\_Covid19}}} & 58.87      &   58.95    & 59.78       & 61.55     &    62.25   &   63.77    &  {\bf{66.50}}       \\
			\midrule
			\multicolumn{3}{c|}{{\textit{Twitter15$\rightarrow$Twitter16}}} &  75.15     &   75.17    &   76.32    &  77.23     &   77.89    &   78.23    &   {\bf{79.47}}     \\
			\midrule
			\multicolumn{3}{c|}{{\textit{Weibo$\rightarrow$Weibo\_Covid19}}} &  57.96     &  58.16     & 60.03      &  63.16     &   65.23    &  65.79     &   {\bf{68.92}}   \\
			\bottomrule
		\end{tabular}%
            \caption{Ablation study results (\%) on four groups of cross-domain datasets }
		\label{tab3}%
	\end{table*}%
	

	\subsection{Overall Performance}
	Table \ref{tab2} shows the performance of the UCD-RD method and all the compared methods on the four groups of cross-domain datasets. From Table \ref{tab2}, the first group of experiments is based on the time series rumor detection methods, and the second group of results is based on the propagation structure rumor detection approaches. We can observe that the results of the first group of experiments are worse than those of the second group of experiments, which proves the propagation structure-based methods also excel the time series-based methods on the cross-domain datasets.
	
	Among the propagation structure-based baselines in the second group, since these methods mainly focus on learning the rumor representation for in-domain rumor detection, the performance is worse on the cross-domain datasets due to the domain shift. Different from that, we propose the UDA-RD method to alleviate the domain shift by aligning the feature representation between a pair of source and target domains. According to Table \ref{tab2}, we found that UCD-CEloss gets better performance compared with these baseline models, which proves that using the self-attention mechanism to model the path embeddings of rumors enables to learn the discriminative rumor patterns. On the basis of UCD-CEloss, we add the instance-wise and prototype-wise contrastive learning and cross-attention mechanism, and the performance has significant improvement, especially on the {\textit{Terrorist$\rightarrow$Gossip}} cross-domain dataset, which achieves a performance improvement of 14.69\% on accuracy. It demonstrates that our method can alleviate the problem of domain shift and gets state-of-the-art performance.
	
	
	
	\subsection{Ablation Study}
	We investigate the effectiveness of each loss function in UCD-RD on four groups of cross-domain datasets. Table \ref{tab3} shows that adding each component contributes to the final results without any performance degradation. 
	From Table \ref{tab3}, we can find that the cross-domain contrastive learning and cross-attention module play important roles in our model. The cross-domain contrastive learning can decrease the discrepancy between a source domain and a target domain, due to the instance-wise and prototype-wise contrastive learning making the representations of the source data and the target data from the same class be closer while keeping representations from different classes far away. Meanwhile, the cross-attention module makes our model more robust and alleviates the noise impact of pseudo labels in the target domain.

	\subsection{Effects of Hyper-parameters}
\begin{figure}[H]
		\centering
		\subfloat[$\alpha$.]{\includegraphics[width=1.62in]{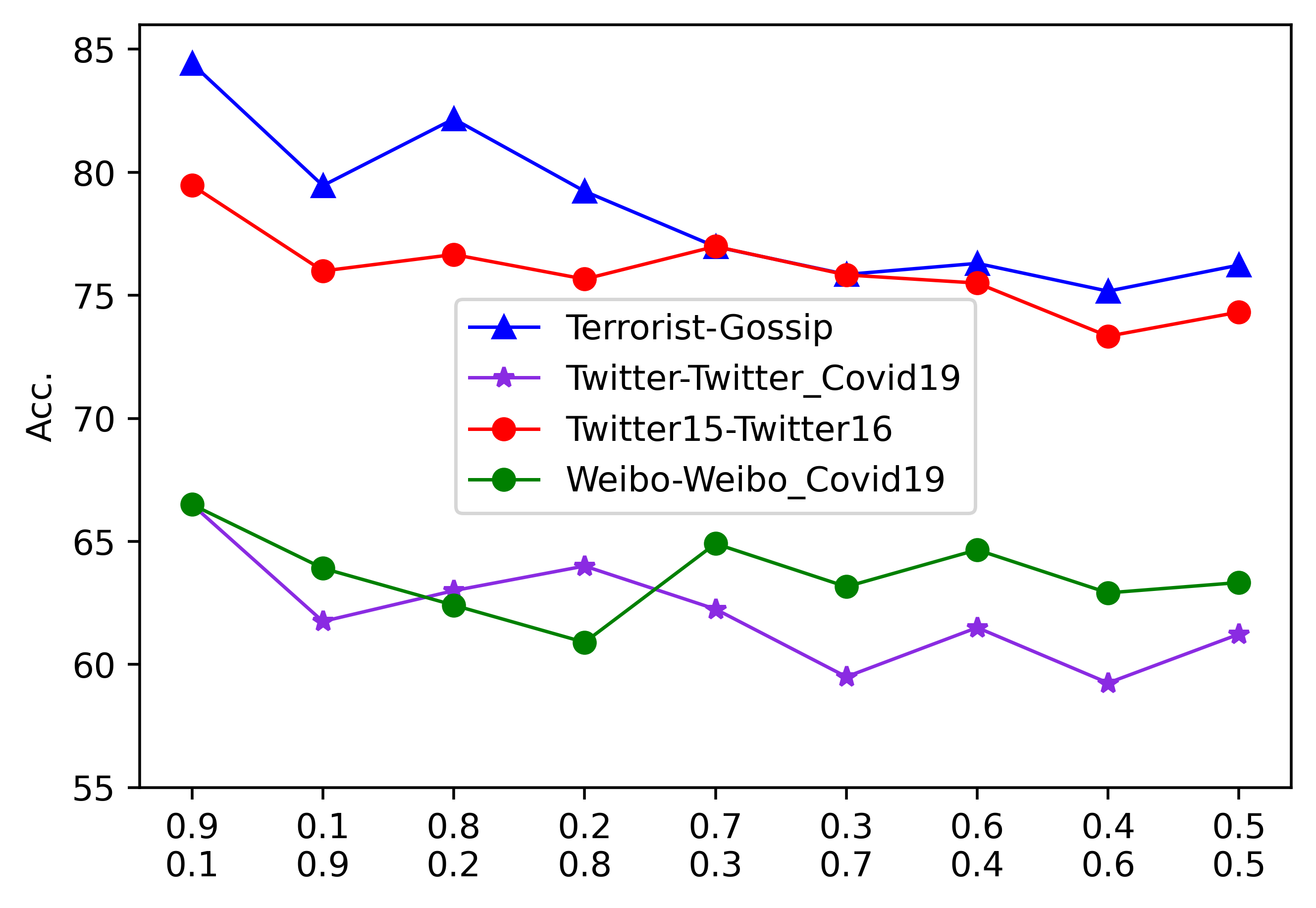}
			\label{alpha}}
		\subfloat[$\beta$.]{\includegraphics[width=1.62in]{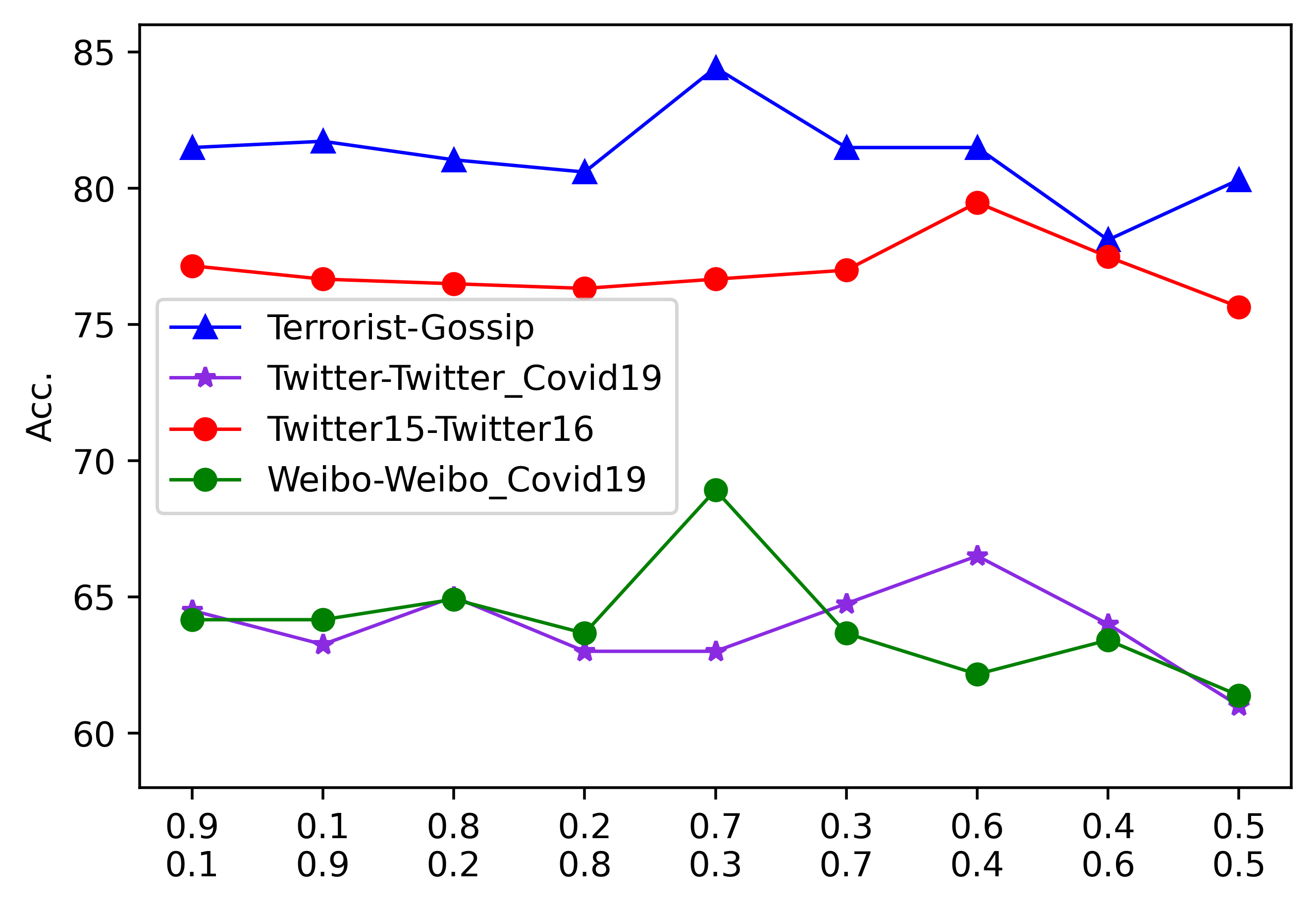}
			\label{beta}}
		\hfil
		\subfloat[$\gamma$.]{\includegraphics[width=3.3in]{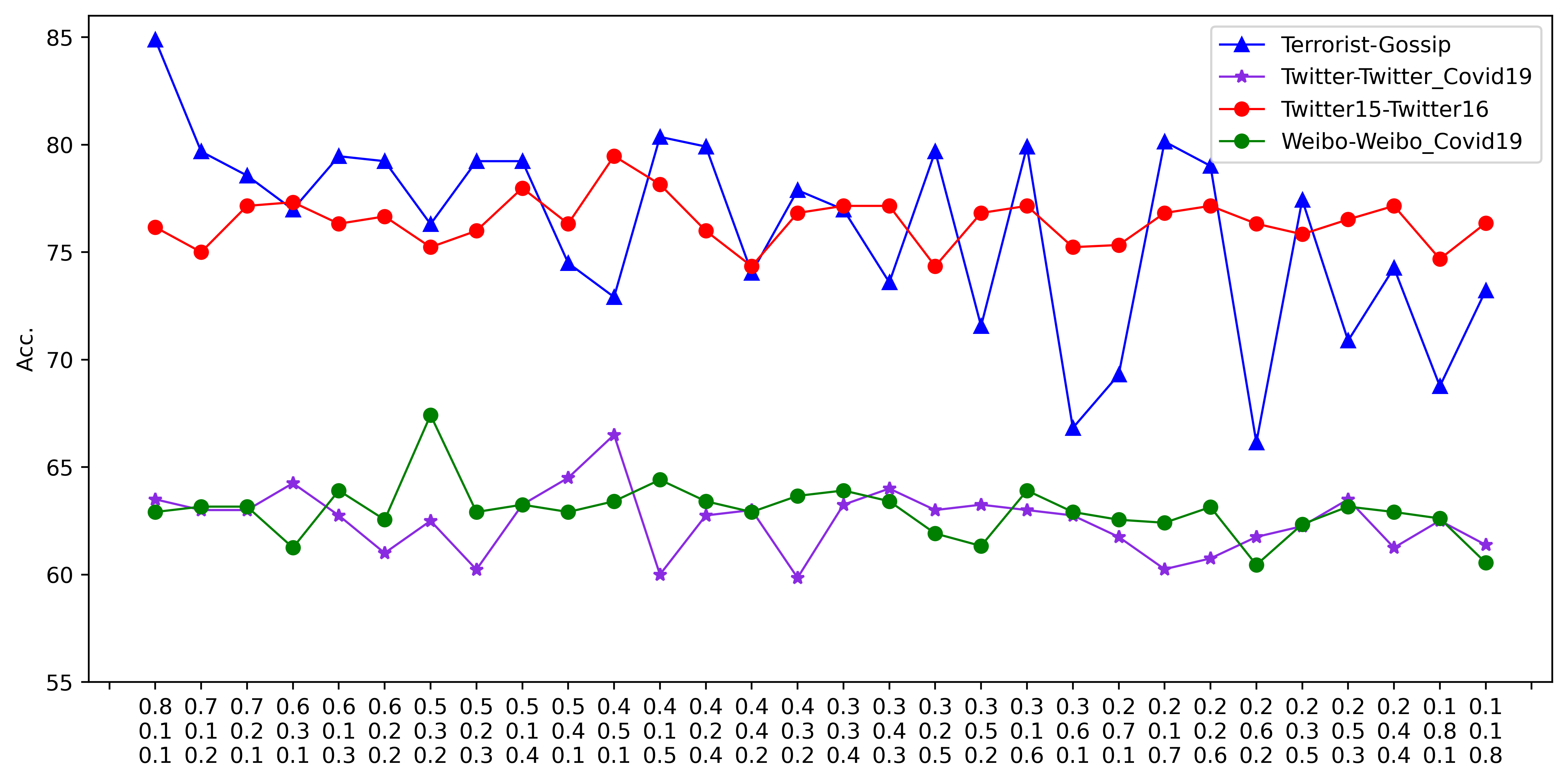}
			\label{gamma}}
		\caption{Performance sensitivity of hyper-parameters $\alpha$, $\beta$, $\gamma$ in UCD-RD.}
		\label{para}
	\end{figure}

	We test the sensitivity of UCD-RD to the $\alpha$, $\beta$, and $\gamma$ on the four groups of cross-domain datasets. As shown in Figure \ref{para}, since the hyper-parameters of $\alpha, \beta, \gamma$ are limited to $\sum\nolimits_{i}{\alpha_i}=1$, $\sum\nolimits_{i}{\beta_i}=1$ and $\sum\nolimits_{i}{\gamma_i}=1$, $\alpha_1, \alpha_2$, $\beta_1, \beta_2$ and $\gamma_1, \gamma_2, \gamma_3$ are appeared in groups in our experiments respectively. Taking Figure \ref{alpha} as an example, each group of values in the x-axis denotes a group of $\alpha_1, \alpha_2$ from top to bottom with fixed $\beta, \gamma$ at their optimal values. Figure \ref{beta} and Figure \ref{gamma} are set similarly. From Figure \ref{para}, we observe that UCD-RD has different sensitivity in different datasets. For instance, for the {\textit{Terrorist$\rightarrow$Gossip}} data, when these hyper parameters $\alpha_1=0.9$, $\alpha_2=0.1$, $\beta_1=0.7$, $\beta_2=0.3$, and $\gamma_1=0.8$, $\gamma_2=0.1$, $\gamma_3=0.1$, UCD-RD achieves the best performance.
    Whereas for the other datasets, the optimal hyper-parameters are at different settings. 
	
	\section{Conclusion}
	We propose a contrastive learning and cross-attention model for cross-domain rumor detection. We build instance-wise and prototype-wise contrastive learning to align features so as to reduce the domain discrepancy between  source data and target data such that the representations of the source data and the target data from the same class are close to each other while those from different classes are far away. Moreover, we use a cross-attention mechanism on a pair of source data and target data with the same labels to learn the domain-invariant features and alleviate the noise impact of pseudo labels. 
   Experiments on four groups of public datasets show that UCD-RD achieves the best performance. 

\section{Acknowledgments}
The authors would like to thank all the anonymous reviewers for their
help and insightful comments. This work is supported in part by the National Natural Science Foundation of China (61876016), the National Key R\&D
Program of China (2018AAA0100302) and Baidu Pinecone Program.

\bibliography{aaai23}

\end{document}